\documentclass[floatfix,aps,superscriptaddress,prd,amsmath,nofootinbib,preprintnumbers,twocolumn]{revtex4-1}

\usepackage{graphicx}
\usepackage{bm}
\usepackage[
colorlinks=true,        
citecolor=blue,         
linkcolor=blue,         
urlcolor=blue           
]{hyperref}             

\newcommand{\nc}{\newcommand*} 
\nc{\Om}{\Omega}
\nc{\ogw}{\Omega_{\mathrm{GW}}}
\nc{\rd}{\mathrm{d}}
\nc{\eg}{\textit{e.g.~}}
\nc{\red}[1]{\textcolor{red}{#1}}
\nc{\lvc}{LIGO/Virgo} 

\def\({\left(}
\def\){\right)}
\def\[{\left[}
\def\]{\right]}

\def\e{\begin{equation}}
\def\q{\end{equation}}
\def\m{\begin{eqnarray}}
\def\n{\end{eqnarray}}

\begin{document}

\title{Measuring the tilt of primordial gravitational-wave power spectrum from observations}

\author{Jun Li}
\email{lijun@itp.ac.cn} 
\affiliation{CAS Key Laboratory of Theoretical Physics, 
    Institute of Theoretical Physics, Chinese Academy of Sciences,
    Beijing 100190, China}
\affiliation{School of Physical Sciences, 
    University of Chinese Academy of Sciences, \\
    No. 19A Yuquan Road, Beijing 100049, China}

\author{Zu-Cheng Chen}
\email{chenzucheng@itp.ac.cn} 
\affiliation{CAS Key Laboratory of Theoretical Physics, 
    Institute of Theoretical Physics, Chinese Academy of Sciences,
    Beijing 100190, China}
\affiliation{School of Physical Sciences, 
    University of Chinese Academy of Sciences, \\
    No. 19A Yuquan Road, Beijing 100049, China}

\author{Qing-Guo Huang}
\email{huangqg@itp.ac.cn}
\affiliation{CAS Key Laboratory of Theoretical Physics, 
    Institute of Theoretical Physics, Chinese Academy of Sciences,
    Beijing 100190, China}
\affiliation{School of Physical Sciences, 
    University of Chinese Academy of Sciences, \\
    No. 19A Yuquan Road, Beijing 100049, China}
\affiliation{Synergetic Innovation Center for Quantum Effects and Applications, 
    Hunan Normal University, Changsha 410081, China}

\date{\today}

\begin{abstract}

Primordial gravitational waves generated during inflation lead to the B-mode polarization in the cosmic microwave background and a stochastic gravitational wave background in the Universe. We will explore the current constraint on the tilt of primordial gravitational-wave spectrum, and forecast how the future observations can improve the current constraint.


\end{abstract}

\pacs{???}

\maketitle


The primordial gravitational waves (GWs) encode the information about inflation (see a recent brief summary in \cite{Li:2019efi}), and the power spectrum of primordial GWs have attracted a lot of attentions. There are in general two crucial properties for the primordial GW power spectrum, namely the amplitude and the tilt $(n_t)$. Even though the B-mode polarization of cosmic microwave background (CMB) can be used to constrain the tensor tilt \cite{Huang:2015gca}, the cosmic variance places an inevitable measuring uncertainty of tensor tilt, i.e. $\sigma_{n_t}=1.1\times 10^{-2}$ for $\ell_{\rm{max}}=300$ in \cite{Huang:2017gmp}. 

In fact, not only do the primordial GWs lead to the B-mode polarization of CMB, but also generate a stochastic gravitational wave background (SGWB) covering very wide frequency bands. After LIGO Science Collaboration announced the first direct detection of GW from the coalescence of binary black holes \cite{Abbott:2016blz}, many experiments are prepared to measure GW in a wide range of frequencies, such as which includes updated LIGO detector, LISA detector, Pulsar timing array (PTA) and so on. All of these observations are sensitive to the stochastic gravitational wave background. 


In order to achieve a better constraint on the tensor tilt, ones should combine observational datasets at different frequency bands. Roughly speaking, the CMB B-mode polarization can constrain the spectrum in the very low frequency band ($10^{-20} \sim 10^{-15}$ Hz), and the most sensitive frequencies for LIGO/Virgo, LISA and PTA experiments are at around $10$Hz, $10^{-3}$Hz and $10^{-8}$Hz, respectively. One can expect to combine these experiments to obtain a much better constraint on the tensor tilt. Actually, the constraint on the positive part of tensor tilt can be significantly improved by combining CMB B-mode polarization data with the LIGO upper limit on the intensity of SGWB in \cite{Huang:2015gka}.

In this letter, we combine the CMB B-mode data from BICEP2 and Keck array through 2015 reason \cite{Ade:2018gkx} and 
the null search results of the SGWB from LIGO O1 and O2 \cite{TheLIGOScientific:2016dpb, LIGOScientific:2019vic} 
to obtain the latest constraint on the tensor tilt. Furthermore, we will also forecast how much the future GW experiments, including LISA, IPTA, and FAST, can improve the constraint on the tensor tilt by combining with CMB B-mode polarization data.

The B-mode component of CMB polarization mainly comes from the tensor perturbation on very large scales and encodes the information about primordial GWs \cite{Kamionkowski:2015yta}. In addition, the primordial GWs also generate an irreducible background. SGWB is a type of gravitational wave produced by an extremely large number of weak, independent and unresolved GW sources. 
It is useful to characterize the spectral properties of SGWB by introducing how the energy is distributed in frequency as follows 
\e
\Omega_{gw}(f)=\frac{1}{\rho_c}\frac{d\rho_{gw}}{d\ln f}=\frac{2\pi^2}{3H_0^2}f^3 S_h(f).
\q
The fractional contribution of the energy density in GWs to total energy density is a dimensionless quantity and $S_h(f)$ is the strain power spectral density of a SGWB. 
For simplicity, the power spectrum of the tensor perturbations is parameterized by
\m
P_t(k)&=&A_t\(\frac{k}{k_*}\)^{n_t},\label{eqs:spectrumtensor}
\n
where $A_t$ is the tensor amplitude at the pivot scale $k_*=0.01$ Mpc$^{-1}$ and $n_t$ is the tensor tilt. If the amplitude of power spectrum decreases with increasing frequency, the spectrum is red-tilted, and if the amplitude grows with the increasing frequency, the spectrum is blue-tilted. For convenience, we introduce a new parameter, namely the tensor-to-scalar ratio $r$, to quantify the tensor amplitude compared to the scalar amplitude $A_s$ at the pivot scale:
\e
r\equiv\frac{A_t}{A_s}.
\q
And then today's GW fractional energy density per logarithmic wave-number interval (the amplitude of this irreducible background) is given by, \cite{Zhao:2013bba,Huang:2015gka}, 
\m
\Omega_{gw}\simeq\frac{15}{16}\frac{\Omega_m^2 A_s r}{H_0^2\eta_0^4k_{\mathrm{eq}}^2}\Big(\frac{k}{k_*}\Big)^{n_t},
\n
where $\Omega_m$ is matter density, $H_0$ is Hubble constant, $\eta_0=1.41\times 10^4$ Mpc denotes the conformal time today and $k_{\mathrm{eq}}=0.073\Omega_mh^2\ \mathrm{Mpc}^{-1}$ denotes the wavenumber when matter-radiation equality.

To characterize the detection ability for a GW detector, it is necessary to calculate
the corresponding signal-to-noise ratio (SNR).
For Advanced LIGO detectors, the SNR is given by \cite{Thrane:2013oya}
\e
\rho= \sqrt{2T} \[\int df \sum_{I, J}^{M} \frac{\Gamma_{IJ}^2(f)S_h^2(f)}{P_{nI}(f)P_{nJ}(f)}\]^{1/2},
\q
where $T$ is the observation time, $P_{nI}$ and $P_{nJ}$ are the auto power spectral densities for noise in detectors $I$ and $J$. 
For an autocorrelation measurement in the LISA detector, 
the SNR can be calculated by \cite{Thrane:2013oya,Caprini:2015zlo}
\e
\rho = \sqrt{T}\[\int df \Big(\frac{\Omega_{gw}}{\Omega_n}\Big)^2\]^{1/2},
\q
where $\Omega_n$ is related to the  strain noise power spectral density $S_n$ by
\e
\Omega_n = \frac{2\pi^2}{3H_0^2} f^3 S_n.
\q
For a PTA measurement, we assume all pulsars have identical white timing noise PSD \cite{Thrane:2013oya}
\e
    S_n = 24 \pi^2 f^2 \Delta t\, \sigma^2,
\q
where $1/\Delta t$ is the cadence of the measurements and $\sigma$ is the root-mean-square timing noise.
Then the SNR can be obtained by
\e 
\rho = \sqrt{2T} \(\sum_{I, J}^{M} \chi_{IJ}^2\)^{1/2} \[ \int \rd f \(\frac{\ogw(f)}{\Om_n(f) + \ogw(f)}\)^2 \]^{1/2},
\q
where $\chi_{IJ}$ is the Hellings and Downs coefficient for pulsars $I$ and $J$ \cite{Hellings:1983fr}. 
Here, we consider two PTA projects, namely IPTA \cite{Verbiest:2016vem} and FAST \cite{Nan:2011um}, respectively.
We make the same assumptions for these PTAs as were presented in \cite{Kuroda:2015owv}.
The number of pulsars, observation times and timing accuracy for these PTAs
can be found in Table 5 of \cite{Kuroda:2015owv}.

\begin{figure}
    \centering
    \includegraphics[width=8cm]{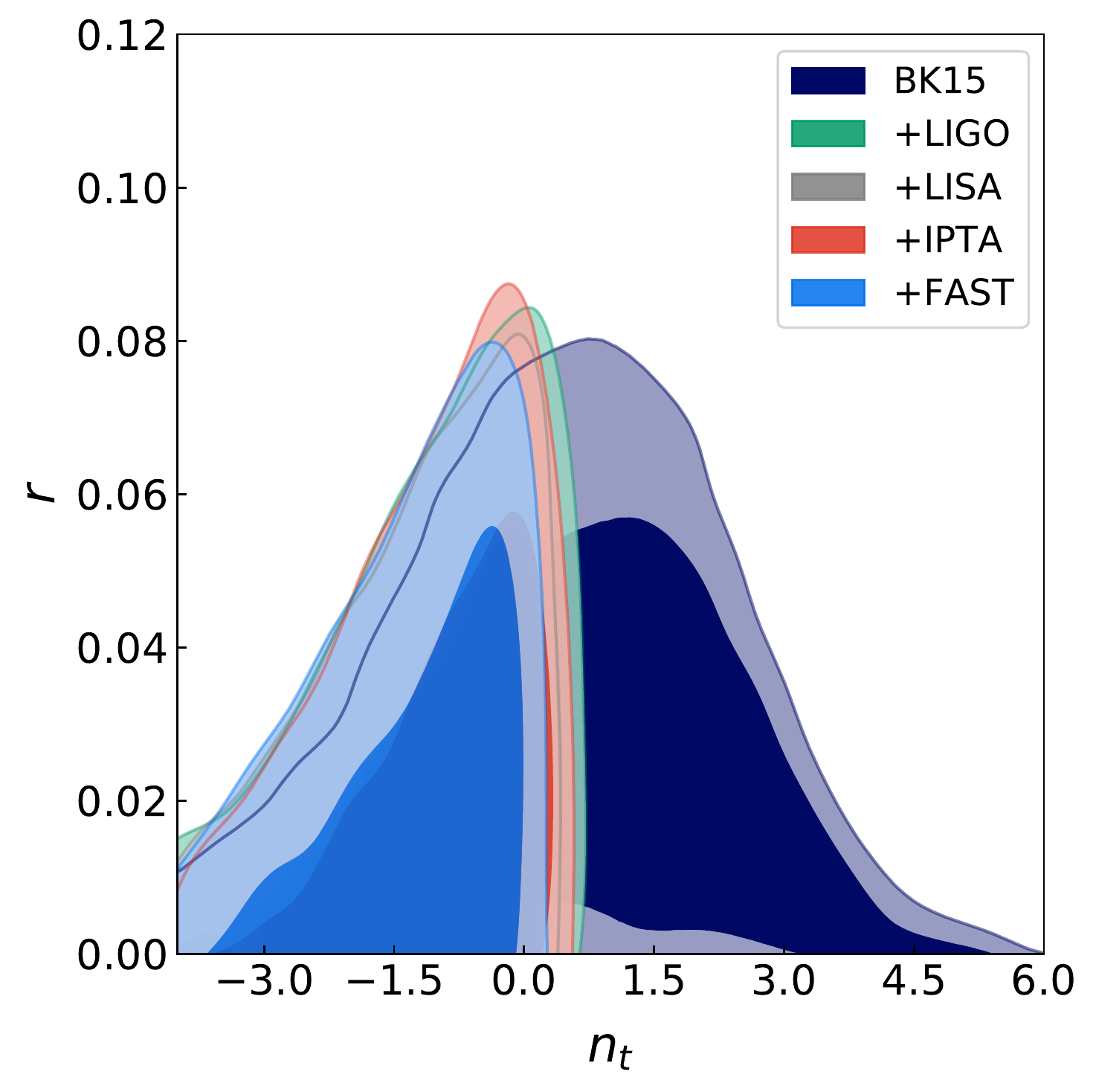}
    \caption{The marginalized contour plot for parameters $n_t$ and $r$ at $68\%\ \mathrm{CL}$ and $95\%\ \mathrm{CL}$ from BK15, BK15+LIGO, BK15+LISA, BK15+IPTA and BK15+FAST datasets, respectively. Here we assume the non-detection of SGWB from future observations including LISA, IPTA and FAST. }
    \label{fig:nt_r}
\end{figure}

\begin{figure}
    \centering
    \includegraphics[width=8cm]{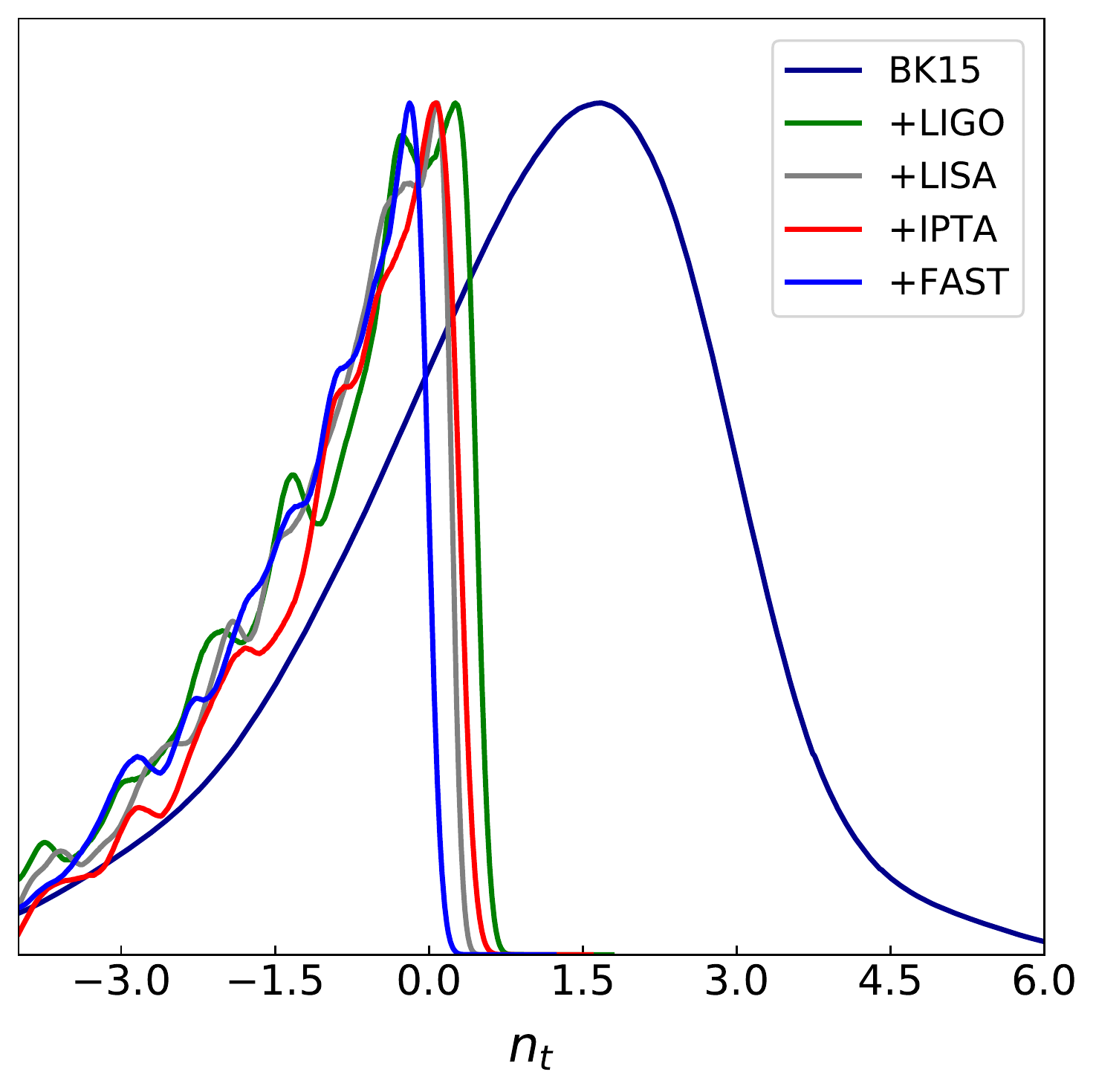}
    \caption{The likelihood distributions of parameters $n_t$ from BK15, BK15+LIGO, BK15+LISA, BK15+IPTA and BK15+FAST datasets, respectively. Here we assume the non-detection of SGWB from future observations including LISA, IPTA and FAST. }
    \label{fig:nt}
\end{figure}

First of all, we adopt the currently available data to constrain the tensor tilt by using the publicly available codes Cosmomc \cite{Lewis:2002ah}. 
Here we take parameters $r$ and $n_t$ as fully free parameters, i.e.~$r\in[0, 1]$, $n_t\in[-4, 6]$, and fix the standard $\Lambda$CDM parameters preferred by Planck observations in \cite{Aghanim:2018eyx}: $\Omega_bh^2=0.02242$, $\Omega_ch^2=0.11933$, $100\theta_{\mathrm{MC}}=1.04101$, $\tau=0.0561$, $\ln\(10^{10}A_s\)=3.047$, $n_s=0.9665$. 
In the $\Lambda$CDM+$r$+$n_t$ model, the constraints on parameters $r$ and $n_t$ from BK15 datasets are given by 
\m
r &<& 0.066\quad(95\% \ \mathrm{C.L.}),\\
n_t &=&  0.96^{+3.01}_{-3.77}\quad(95\%\ \mathrm{C.L.}).
\n
Combining BK15 with LIGO, the constraints on parameters $r$ and $n_t$ become
\m
r &<& 0.068\quad(95\% \ \mathrm{C.L.}),\\
n_t &=& -0.84^{+1.64}_{-2.30}\quad(95\%\ \mathrm{C.L.}).
\n
A scale-invariant primordial GW power spectrum is consistent with the current datasets, and the constraint on the positive part of tensor tilt is significantly improved once the LIGO data is taken into account. See the results in Figs.~\ref{fig:nt_r} and \ref{fig:nt}.

Here we are also interested in exploring the abilities of future GW observations, such as LISA, IPTA and FAST, for constraining the tensor tilt. We assume the non-detection of SGWB for LISA, IPTA and FAST, and then see how these data will potentially improve the constraint on the tensor tilt. 
Similarly, the potential constraints on parameters $r$ and $n_t$ at $95\%$ C.L. are
\m
r &<& 0.065\quad(95\% \ \mathrm{C.L.}),\\
n_t &=& -1.02^{+1.57}_{-2.19}\quad(95\%\ \mathrm{C.L.}), 
\n
from from BK15+LISA datasets; 
\m
r &<& 0.067\quad(95\% \ \mathrm{C.L.}),\\
n_t &=& -0.86^{+1.53}_{-2.19}\quad(95\%\ \mathrm{C.L.}), 
\n
from BK15+IPTA datasets; and 
\m
r &<& 0.063\quad(95\% \ \mathrm{C.L.}),\\
n_t &=& -1.03^{+1.41}_{-2.12}\quad(95\%\ \mathrm{C.L.}).
\n
from BK15+FAST datasets, respectively. The results are illustrated in Figs.~\ref{fig:nt_r} and \ref{fig:nt}.

To summarize, we constrain the tensor tilt from CMB polarization experiments and LIGO interferometer observations, and forecast the potential abilities of LISA detector and PTA projects for measuring tensor tilt. We find that LIGO, LISA and PTA can significantly improve the constraints on the tensor tilt if the amplitude of the tensor power spectrum is not too small to be detected. In particular, FAST may provide a much better constraint on the positive part of the tensor tilt, namely $n_t<0.38$ at $95\%$ C.L..


\noindent {\bf Acknowledgments}.
This work is supported by grants from NSFC (grant No. 11690021, 11575271, 11747601), the Strategic Priority Research Program of Chinese Academy of Sciences (Grant No. XDB23000000, XDA15020701), and Top-Notch Young Talents Program of China.
This research has made use of \textit{GWSC.jl} 
(\url{https://github.com/bingining/GWSC.jl}) package to calculate
the SNR for various gravitational-wave detectors.



\end{document}